\documentclass[%
reprint,
superscriptaddress,
amsmath,amssymb,
aps,
]{revtex4-2}

\usepackage{graphicx}
\usepackage{color}
\usepackage[normalem]{ulem}
\usepackage{dcolumn}
\usepackage{bm}
\usepackage{upgreek}
\usepackage{layouts}
\usepackage{makecell}

\begin{document}
	\preprint{APS/123-QED}

	\title{Path-Degenerate Quantum Interferometry for Loss Resilience in \\Gravitational-Wave Detectors}

	\author{Jonas Rittmeyer}
    \author{Niels B\"ottner}
    	\affiliation{%
		Institut f\"ur Quantenphysik und Zentrum f\"ur Optische Quantentechnologien\\
		Universit\"at Hamburg, Luruper Chaussee 149, 22761 Hamburg, Germany}
    \author{Farid Khalili}
        \affiliation{Russian Quantum Center, Skolkovo IC, Bolshoi Boulevard 30, Building 1, Moscow, 121205, Russia}
        \affiliation{Faculty of Physics, M V Lomonosov Moscow State University, Moscow 119991, Russia}
	\author{Mikhail Korobko}	
	\author{Roman Schnabel}
	\email{roman.schnabel@uni-hamburg.de}
	\affiliation{%
		Institut f\"ur Quantenphysik und Zentrum f\"ur Optische Quantentechnologien\\
		Universit\"at Hamburg, Luruper Chaussee 149, 22761 Hamburg, Germany}

\date{\today}

\begin{abstract} 
Gravitational wave (GW) detectors utilise squeezed vacuum states of light to reduce quantum noise relative to the signal. Optimised squeeze angles quantum-correlate photon shot noise and quantum back-action noise. In this way, sensitivity in the quantum non-demolition (QND) regime is achievable. However, the decoherence caused by loss of photons partially destroys the quantum correlations, which is a challenge even in future, highly optimised GW detectors such as the Einstein Telescope. With higher loss resilience, the QND performance could be enhanced even further by the `variational output' technique, which optimizes the measured field quadrature for every signal frequency. So far, the level of loss and experimental complexity prevented it for being considered for future GW detectors. Here we propose and partially experimentally demonstrate the path-degenerate quantum interferometry scheme that promises lower photon loss due to fewer optical components, mitigates the decoherence effect caused by loss in the output channel, reduces the impact of spatial mismatches between different optical modes and automatically integrates a variational output scheme without additional components. Our proposal delivers a further improved signal-to-quantum-noise ratio and has the potential to enhance the sensitivity of both existing and planned GW detectors whilst simultaneously reducing manufacturing costs and operational complexity.
\end{abstract}

\maketitle

\section*{Introduction}\vspace{-2mm}
By the late 1950s, it had become widely accepted that gravitational waves (GWs) can release a tiny amount of their energy and are therefore, in principle, measurable~\cite{Preskill1995}. The strongest signals were predicted to come from astrophysical events \cite{Ostriker1969,Press1971,Press1972}, however, they are so weak when finally reaching GW detectors, that quantum mechanical uncertainties in the measurement process can easily mask them~\cite{Glauber1963,Weiss1972}. 
In all current GW detectors, test masses of spacetime are continuously read out with quasi-monochromatic laser light. Along with quantum measurement (shot) noise, quantum back-action (radiation pressure) noise constitutes the fundamental sensitivity limitation~\cite{Braginsky1968,Caves1981}.
Quantum non-demolition (QND) measurements use engineered quantum correlations in the measurement process to avoid the impact of back-action noise~\cite{Braginsky1980}. 
One QND approach optimizes the squeeze angle~\cite{Unruh1983,Jaekel1990} via a filter cavity (`frequency-dependent squeezed input') \cite{Kimble2001,Chelkowski2005,Khalili2010a,Barsotti2019} of the squeezed vacuum states incident on the signal-output port~\cite{Caves1981,Schnabel2010,Vahlbruch2010,LSC2011,Grote2013,Schnabel2017}. 
Another QND approach (`variational output')~\cite{Kimble2001, Khalili2007} exploits quantum correlations in the test-mass reflected light created by the radiation pressure uncertainty, by selecting the optimal measurement quadrature. 
Recently, at LIGO, frequency-dependent squeezed input allowed to achieve broadband quantum noise reduction reaching up to 3\,dB into the QND regime (inferred after subtraction of technical noises)~\cite{Jia2024}.
The same concept is also planned for future GW detectors, such as Einstein Telescope~\cite{Punturo2010, korobko2025quantum} and Cosmic Explorer~\cite{Evans2021}. 
These detectors are expected to achieve a total strain sensitivity in the QND regime over part of their observation band.
To achieve this goal, it is necessary not only to significantly reduce classic noise sources, but also to minimise quantum decoherence in squeezed light, which has so far prevented the attainment of high squeezing values and the introduction of variational output.
Relevant decoherence channels can be separated into three major categories: i) internal loss of the interferometer itself; ii) input loss: e.g.~from Faraday rotators~\cite{Genin2018}, which so far are required to break degeneracy of input and output along the signal channel~\cite{Kimble2001, McKenzie2002} or from filter cavities, including mode mismatch effects\,\cite{Kwee2014,McCuller2021}; and iii) output loss: e.g.~from the output mode cleaner~\cite{Capote2025} and the imperfect quantum efficiency of the balanced homodyne detector (BHD), including that of the photodiodes, depending on the laser wavelength~\cite{Vahlbruch2016,Darsow-Fromm2021,Meylahn2022,Albers2026}.
Circumventing the output losses mentioned above can be largely done by utilising parametric amplification either externally~\cite{Caves1981, Manceau2017, Knyazev2019, Frascella2021, Salykina2023, Kwan2026} or internally \cite{Korobko2017, korobko2023, korobko2023fundamental, Vermeulen2026}. But until now either would require a significant redesign of the GW detector or an overall increase of experimental complexity. 
  
\begin{figure}
\includegraphics[width=1\linewidth]{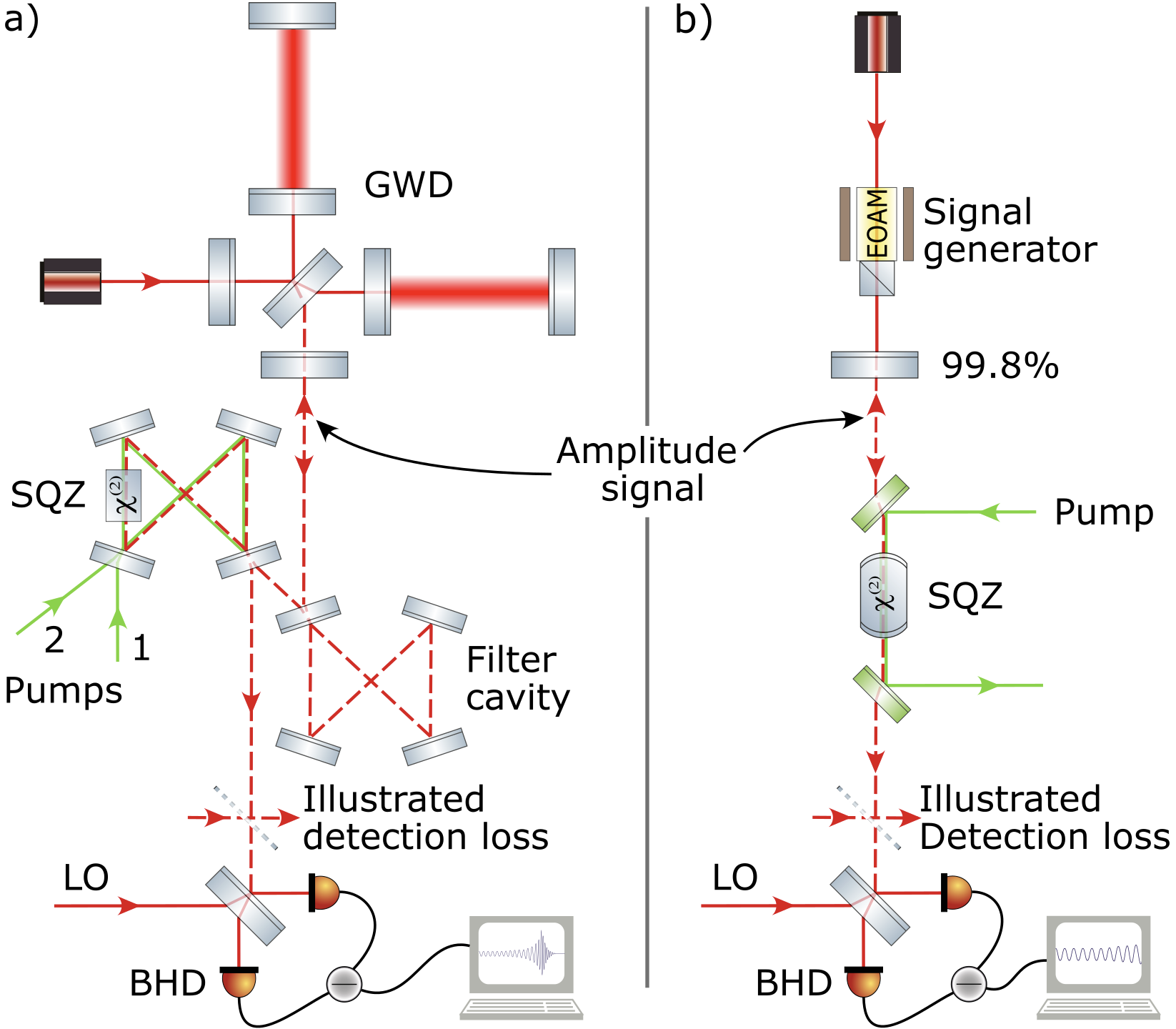}
\caption{
{\bf Path-degenerate quantum interferometry} --- (a) Our proposal for GW observatories. Different from current instruments, input and output optical axes between interferometer and photo-electric detection (balanced homodyne detector, BHD) are degenerate, and squeeze resonator (SQZ) and detuned filter cavity are used twice. 
A Faraday rotator is not required. Filtering implements `variational input' and `variational output' \cite{Kimble2001}, and the second pass through the squeeze resonator is used for noiseless amplification. 
(b) Our simplified experimental realisation, emulating the interferometer and its signal in the shot-noise dominated regime (no filter cavity required) by a retro reflector and a monochromatic amplitude modulator. We use a double-pass squeezer that is singly-resonant at the pump wavelength, which symmetrically squeezes and amplifies the input and output, respectively. LO: local oscillator; EOAM: electro-optic amplitude modulator.
}
\label{fig:1}
\end{figure}

Here, we propose a novel quantum interferometry scheme for GW detectors, which is path-degenerate between the interferometer and the photo-electric detection. The optical axes of input and output beams are not separated using Faraday optics, and squeeze resonator and filter cavity are used twice in opposite directions, see Fig.\,\ref{fig:1}. We demonstrate the scheme in the shot-noise dominated regime in a proof-of-principle experiment. 
Our proposal offers three advantages: it does not require Faraday optics in the signal path and therefore potentially exhibits lower input and output losses; it is robust against measurement losses owing to parametric amplification of the output signal without additional components; and it utilises a variant of the variational output technique \cite{Kimble2001, Khalili2007}.  
Furthermore, our proposal has the potential to significantly simplify the process of mode matching between the output mode cleaner~\cite{Capote2025} (not shown in Fig.\,\ref{fig:1}) and the interferometer, thereby improving the results and further reducing losses.

\section*{Concept and sensitivity}\vspace{-2mm}
Quantum back-action (QBA) turns a vacuum state entering an interferometer with movable mirrors into a squeezed state through the process called ponderomotive squeezing\,\cite{Braginsky1967,Safavi-Naeini2013,Purdy2013}. 
The outgoing state is squeezed in different phase angles at different frequency, which manifests as anti-squeezed (quantum radiation pressure) noise at low frequency.
When the incoming vacuum state is replaced by an externally produced squeezed state~\cite{Caves1981, LSC2011, Tse2019, Acernese2019}, ponderomotive squeezing acts similarly on it. In order to counteract this effect, the injected squeezed state needs to be produced with a frequency dependent squeeze angle. Such ``frequency-dependent squeezing'' is achieved via reflection of the initially frequency-independent squeezed state from a detuned filter cavity~\cite{Kimble2001,Chelkowski2005,Jia2024}. 
Until now, the injected squeezed modes and the back-reflected signal-carrying modes have been specifically routed along separate paths using Faraday optics. This in principle allows to introduce detuned filter cavities for an optimal frequency-dependent phase rotation of the externally and ponderomotively squeezed GW-signal with respect to the BHD's local oscillator, which was named ``squeezed-variational interferometer''~\cite{Kimble2001}. 

However, the Faraday rotator introduces additional optical loss, and the difficult-to-adjust mode matching -- from the squeeze resonator through  the filter cavities and the resonator-enhanced interferometer towards the output mode cleaner and the BHD -- may lead to hyperloss~\cite{McCuller2021, Grebien2026, Kuns2026}. 
To circumvent the losses at the output mode cleaner and the BHD (here: the ‘measurement losses’), optically parametric amplification has been proposed~\cite{Caves1981}. With perfect coupling to an ideal amplifier, measurement losses can be compensated for to zero, even if the photodiodes have poor efficiency~\cite{Manceau2017, Knyazev2019, Frascella2021, Kwan2026}. 

In the concept proposed here, the degeneracy of the optical axes of the incoming and outgoing modes is retained between the photoelectric detection and the interferometer. Consequently, all optical components along this path are traversed twice, once in each direction. This includes both the filter cavity, if present, and the squeezer cavity itself. The Faraday isolator for squeezed light injection is not required any more, and no additional components are needed compared to the current LIGO and Virgo implementations.
On the forward path, a vacuum state entering the squeezing cavity is squeezed along the direction defined by the phase of the frequency-doubled pump. Then the created state is reflected off the detuned filter cavity, acquiring the necessary frequency dependence. On the backward path the state is reflected off the same filter cavity again, rotating the signal quadrature in a frequency-dependent way. Then it passes through the squeezer again in the opposite direction, fully decoupled from the forward path. A second frequency-doubled pump selects an opposite phase, amplifying the outgoing state instead of squeezing it. As a result, both signal and noise are proportionally increased far above the shot noise level, and any down-stream loss leading to mixing of vacuum state becomes irrelevant.
\\
\begin{figure*}
\includegraphics[width=1\linewidth]{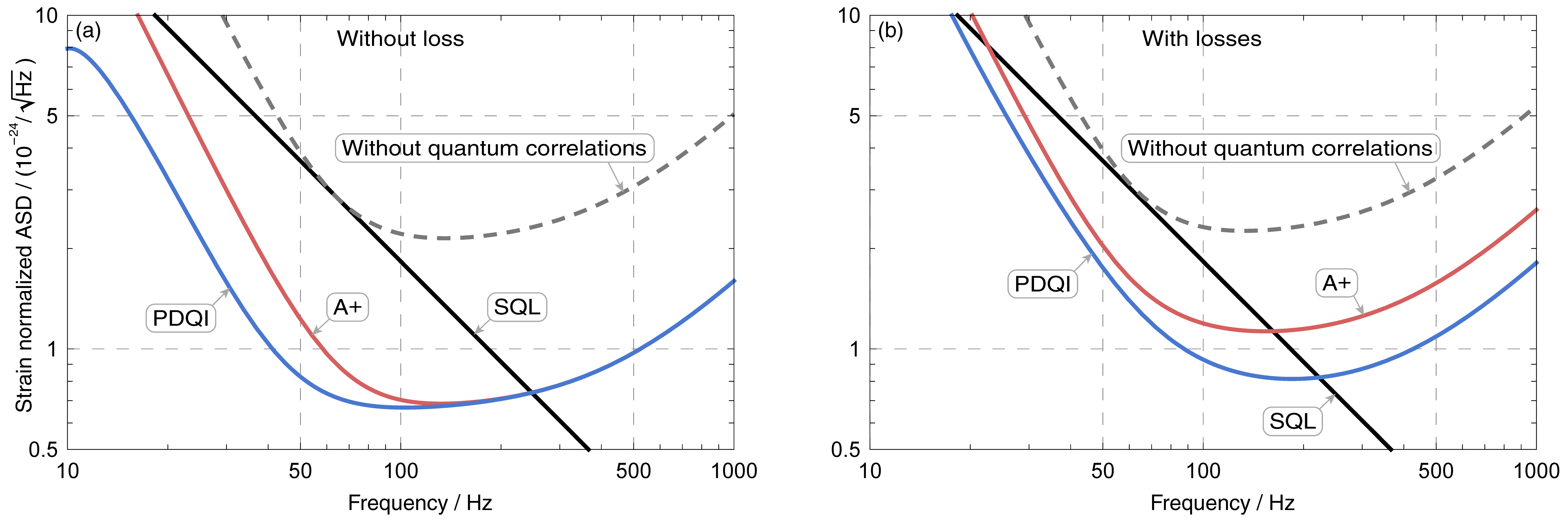}
\caption{
{\bf Quantum noise spectral densities} --- 
Analytically calculated quantum measurement noise and back-action noise of the current design of A+ (with two Faraday rotators and a 300-m standing-wave filter cavity) and PDQI according to Fig.\,\ref{fig:1} with a 300-m Bowtie filter cavity using the A+ parameters in Ref.\,\cite{Barsotti2018}. (a) Without decoherence; 10-dB frequency-dependent input squeezing and filter cavity linewidths of $\gamma_{\text{PDQI}}=31.57\,\text{Hz}$ and $\gamma_{\text{A+}}=44.97\,\text{Hz}$, with optimized detunings, respectively (see tables in the supplementary material). At low frequencies, the positive effect of rotating the readout quadrature in analogy to variational output is discernible. 
(b) With filter-cavity round-trip-losses of $61\,\text{ppm}$ and $123\,\text{ppm}$, for A+ and PDQI respectively, measurement losses of 10\% and an additional loss of 1\% per Faraday rotator in case of A+. Here, the optimal filter cavity linewidths read $\gamma_{\text{PDQI}}=35.55\,\text{Hz}$ and $\gamma_{\text{A+}}=50.14\,\text{Hz}$. The detunings are optimized accordingly. At high frequencies, the resilience to measurement loss is evident. 
In the case of quantum noise dominance, the PDQI curve represents an improved average detection rate by a factor of $2.8$ ($1.41^3$). SQL: Standard quantum limit, below which measurements in the QND-regime are realized if quantum noise dominates.
}
\label{fig:2}
\end{figure*}
Fig.\,\ref{fig:2} compares the spectral densities of the quantum noise terms in the currently prepared Advanced LIGO successor A+ \cite{Barsotti2018}, both in the current implementation and following the `path-degenerate quantum interferometer' (PDQI) proposed here. Fig.\,\ref{fig:2}\,(a) compares the ideal cases, i.e.~without any decoherence. Low-frequency sensitivity is increased compared to the conventional frequency-dependent squeezing (FDS) due to an additional rotation of the readout quadrature on the backward path. Fig.\,\ref{fig:2}\,(b) includes realistic decoherence sources in the interferometer, injection and measurement. We can see the PDQI's resilience against measurement loss $l$ by comparing the signal-to-noise ratio (SNR) for PDQI and FDS limited by shot noise:
\begin{equation} \label{eq:loss-resilience}
    \dfrac{\text{SNR}_{\text{PDQI}}}{\text{SNR}_{\text{FDS}}}= 1 + l (\beta -1),
\end{equation}\noindent
which comes from the parametric amplification strength $\beta$. For non-zero measurement loss, any amplification above 1 results in improved SNR. Our approach can be applied to the existing as well as planned detectors, such as Einstein Telescope and Cosmic Explorer, leading to qualitatively similar sensitivity improvements.

\section*{Proof-of-principle experiment}
\begin{figure} 
\includegraphics[width=1\linewidth]{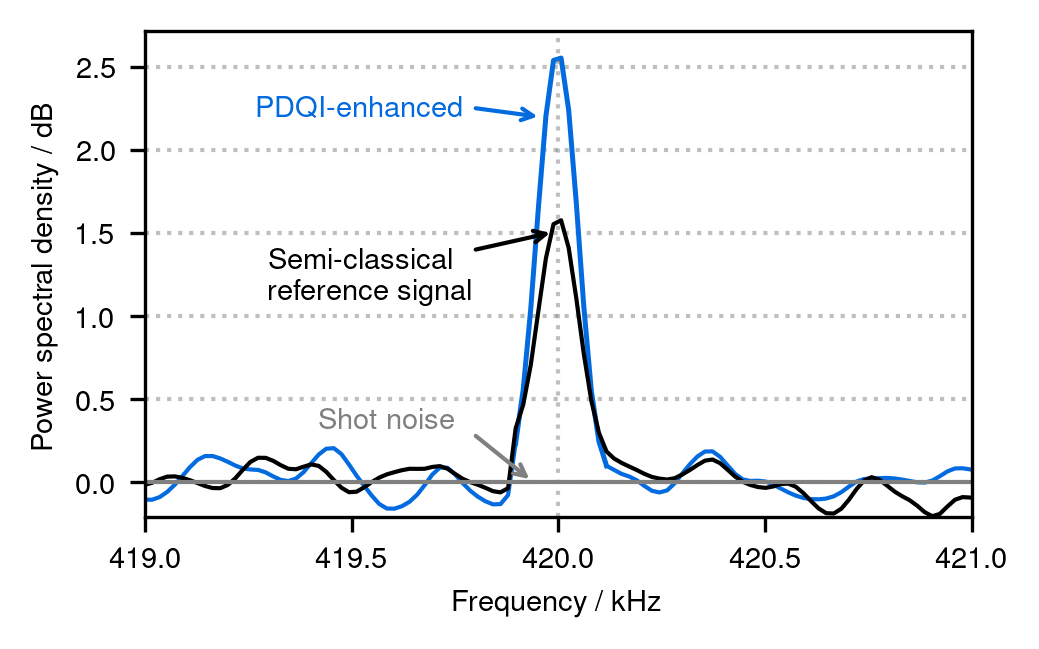}
\caption{
{\bf Shot-noise preserving signal enhancement} --- 
Shown are two shot noise-limited measurement curves for the same input signal amplitude. With PDQI (squeezed input - parametrically amplified output with inverted gain), the signal increased by approximately 1\,dB compared to the semi-classical reference (no parametric gains). The value of 1\,dB was limited by the maximum parametric gain achievable in our single-pass amplifier.
}
\label{fig:3}
\end{figure}

\vspace{-2mm}
Underlining the theoretical proposal of our PDQI scheme we set up an experiment, see Fig.\,\ref{fig:1}\,b), to demonstrate loss resilience of path-degenerate input squeezing combined with output anti-squeezing. As a special case, we further demonstrated shot-noise preserving signal enhancement.\\
In our PDQI setup, the incoming vacuum modes were squeezed, then superimposed with a signal, and the resulting field was then parametrically amplified on the return pass through the same squeezer. Our (two-directional) single-pass squeezer produced a (single-pass) squeeze factor of about 1\,dB, limited by the high pump intensity requirement combined with intensity-dependent imperfections, see the Methods section.
\\
We emulated a signal from an interferometer operated close to a dark port by transmitting a $420\,\text{kHz}$ amplitude modulated carrier field through a retro-reflector, see Fig.\,\ref{fig:1}\,b). 
The readout was done on a BHD, which is also planned for all future ground-based GW detectors. With optimally adjusted phase values the incoming vacuum state got squeezed, and both the noise and signal got parametrically amplified towards detection afterwards. A standing-wave resonator for the pump field resulted in the equal the amplification and squeeze values. After the two interactions the noise remained at the shot noise level, the signal was amplified on the output path, and hence the SNR was increased. This effect is shown in Fig.\,\ref{fig:3} where we observed the shot-noise preserving signal enhancement and the corresponding improvement in the SNR. The semi-classical reference measurement was performed without either input squeezing or output amplification. 
\begin{figure} 
\includegraphics[width=1\linewidth]{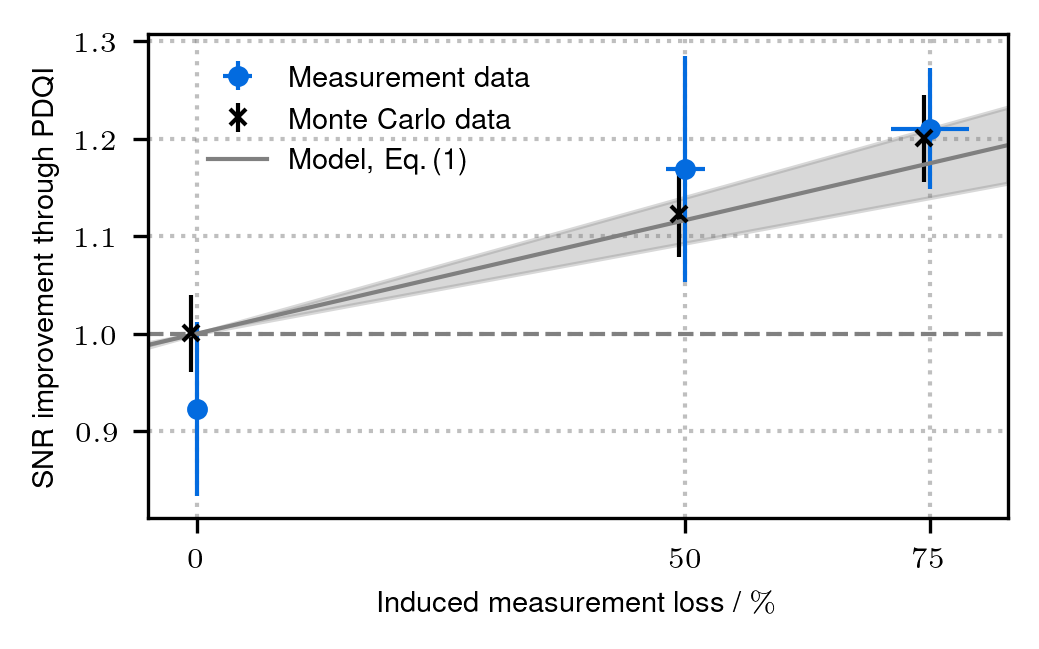}
\caption{
{\bf Loss resilience of PDQI} --- 
The figure shows the improvement in the signal-to-noise ratio (SNR) achieved by PDQI compared with conventional squeezed input ($\approx\!1$\,dB), based on experimental data. To this end, we artificially increased the optical measurement loss following the second pass through the squeezed resonator by
$0$, $50$, and $75\,\%$. Our theoretical analysis used Eq.\,(\ref{eq:loss-resilience}) as a model as well as a Monte-Carlo simulation. Even with our small squeeze factor, a 75\% loss still resulted in a 20\% improvement.
}
\label{fig:4}
\end{figure}

In Fig.\,\ref{fig:4} we demonstrate the loss resilience of PDQI. For that, we compare acquired PDQI data with expected SNR improvement in a traditional readout given our squeezing data. As the measurement loss grows, (here of up to $75\,\%$), PDQI is able to mitigate a significant portion of it, resulting in SNR improvement relative to traditional squeezing. 
The PDQI data in Fig.\,\ref{fig:4} were extracted from several measurements, one of which is shown in Fig. \,\ref{fig:3}. Each calculated gain was then compared to the extrapolated gain of a traditionally squeezed readout with the same squeeze value. This data generally matches the theoretical expectation based on Eq.\,(\ref{eq:loss-resilience}) using a Monte-Carlo simulation to account for model uncertainties, see the Methods section.

\section*{Discussion and conclusion}  \label{sec:discussion} \vspace{-2mm}
Optical loss and mode mismatch is a crucial limitation to quantum advantage in GW detectors~\cite{Grebien2026}. Our PDQI approach holds the promise to significantly lighten the requirements on the optic quality and detector design in general. Already current generation of GW detectors could benefit from adopting PDQI due to high modematching losses at the output mode cleaner~\cite{Capote2025}, which in our design would occur after the amplification, and thus their effect would be diminished.
Another benefit of PDQI is enabling the promising future shift to a longer laser wavelength of around 2\,$\upmu$m \cite{SteinlechnerJ2018,Barsotti2019} where the detection efficiencies of photodiodes are lower \cite{Mansell2018, Gurs2026}, which can be mitigated by output amplification~\cite{Kwan2026}.

The following points must be taken into account when implementing PDQI.
First, the optimal design for the squeeze resonator is a travelling-wave resonator (e.g.~a `bow-tie' resonator), as shown in Fig.\,\ref{fig:1}\,a). Its key property is that it allows the resonance enhancement not only for the pump light, but also for the squeezed light when used twice in the path-degenerate way. This significantly reduces the required pump intensity compared to our experiment and allows for independently adjustable parametric gains for quantum input and output. Separate levels for input and output parametric gain are required as the input squeezing is limited by optical loss and introduced anti-squeezing, while the output amplification should be as strong as possible to achieve\,\cite{Manceau2017b, Manceau2017}.
Second, the filter cavity is passed through twice, which doubles its loss contribution. In a PDQI implementation, the filter cavity has a bow-tie design to naturally separate the incoming and reflected beams, unlike the current standing-wave filter cavity design, where this is achieved with (another) lossy Faraday rotator. Our bow-tie filter cavity avoids this Faraday rotator, but comes with additional mirror reflections and therefore a larger round-trip loss. Relevant for the total loss is, however, the loss per unit length~\cite{Khalili2010a,Evans2013}. 
For a fixed spacing between the input and the back mirrors, the double round-trip length of a bow-tie keeps the loss per unit length the same as in a standing-wave filter cavity. 
However, the coupling of the circulation directions through scattered light can have an adverse effect\,\cite{Evans2013}. The implications must be investigated quantitatively. 
If a standing-wave cavity are used in double-pass, polarization non-degeneracy of the cavity due to birefringence of the mirror coatings could prevent from achieving optimal detuning on two passes and further cavity optimisation is required to minimize the effect. 
Ultimately, our approach can be either directly adapted to existing detectors with standing-wave filter cavities, accepting the Faraday losses and increased polarization complexity, or optimized for future detectors by using bow-tie filter cavities.
Third, the degenerate path makes the spatial-mode matching of the input and output fields reciprocal: in the absence of parasitic effects, matching the squeezed field to a filter cavity and interferometer also matching the returning signal field to the same optical train. PDQI can therefore reduce the number of independently controlled mode-matching interfaces, which are one of the major limitations to quantum enhancement in the detectors~\cite{McCuller2021, korobko2025quantum,Kuns2026}. A quantitative assessment of the resulting reduction in multimode decoherence and hyperloss requires a full higher-order-mode model and further research.

PDQI offers exciting possibilities for previously studied alternative designs of GW detectors. In particular, quantum speedmeter~\cite{Braginsky1990,Chen2003} topology allows to evade a significant part of quantum radiation pressure (QBA) noise and reach into the QND regime using only frequency-independent squeezed light. This removes the need for costly and complex filter cavities. Nonetheless, as any QND approach, it is particularly sensitive to optical losses. Our PDQI removes the need for Faraday optics, improves the modematching losses and mitigates measurement losses. Further study is required to show the detailed implementation of PDQI in a realistic speedmeter configuration.

Beyond GW detection, PDQI promises quantum enhancement in the systems where the path to the sensor has path-degenerate input and output. This differs from the canonical two-stage SU(1,1) interferometer, in which two distinct parametric devices prepare and recombine the quantum state around a separately accessed sensing region~\cite{Yurke1986, Salykina2023}. 
PDQI brings particular benefit to the sensors featuring a Michelson interferometer, such as dark matter searchers and quantum gravity detectors~\cite{Chou2016,Vermeulen2021,Patra2025}, high-frequency gravitation wave detectors\,\cite{Aggarwal2025} and table-top force sensors\,\cite{Kleybolte2020}.

Our table-top experiment demonstrated proof-of-principle resilience of PDQI to measurement losses, maintaining quantum advantage even with 75\% added loss. This demonstrates that a single squeeze resonator, in combination with an interferometer, can be operated by adjusting the pump phase and optical path lengths in such a way that downstream optical losses due to imperfect output mode cleaner and BHD are substantially less limiting, enabling quantum advantage that would be otherwise inaccessible. An additional parametric amplifier is not necessary. We recommend that our PDQI should be considered for A+/A$^\sharp$ and the design of the Einstein Telescope as well as the Cosmic Explorer.
\\

\section*{acknowledgments}\vspace{-3mm}
This work was supported and partly financed (NB) by the DFG under Germany's Excellence Strategy EXC~2121 ``Quantum Universe'' -- 390833306.\\[-1mm]

\section*{Author Contributions}\vspace{-3mm}
JR set up and carried out the experiment. NB performed theoretical calculations and simulations. FK and MK initially proposed to use the same standing-wave squeezer twice to achieve loss resilience. MK and RS proposed to use the filter cavity twice. RS proposed using a bow-tie squeeze resonator for path-degenerate input-output conditions and made his resources at the University of Hamburg available to the project. JR, NB, MK and RS wrote the manuscript and produced the figures.
\\[-5mm]
%


\section*{Methods}\vspace{-3mm}

\subsection*{Experimental}
The experiment used a monolithic biconvex crystal of periodically poled KTP, 11.6\,mm in length. The coatings on both optical surfaces were anti-reflective for 1550\,nm and provided reflectivities of $R_{775,1}=99.8\,\%$ and $R_{775,2}=99.93\,\%$ for the pump light. Anti-reflection for 1550\,nm was necessary to avoid mixing of the two directions. Hence to be able to achieve high squeeze values the reflectivities of the 775\,nm pump had to be high to create strong intensity in the crystal. Yet the crystal produced only 1\,dB of squeezing. This value was defined by the moderate efficiency of the parametric process, which we assume was limited by the mode deformation of the pump light by a thermal lens and/or photo darkening of the crystal. A bow-tie design as used in LIGO\,\cite{Tse2019} would circumvent these problems.
\\
The squeezed vacuum states were superimposed with a 420\,kHz amplitude modulated signal and a weak carrier of $\approx 0.5\,\upmu$W on retro-reflector ($R_{R}=99.8\,\%$ at 1550\,nm), see Fig.\,\ref{fig:1}\,b). The carrier was additionally phase modulated at 5\,MHz to employ a control loop for stabilising the BHD on amplitude modulation measurement. 
For the data acquisition of Fig.\,\ref{fig:3} the phases of the pump ($\varphi_{\text{sqz}}$) and the signal output ($\varphi_{\text{amp}}$) were manually set to preserve shot noise while maximally amplifying the signal. The resulting improvement is given by $G_{\text{PDQI}}= \text{e}^{2r} \approx\,1.25 \;\widehat\approx \,+1\,\text{dB}$ with the squeeze / amplification parameter $r$.
For the data acquisition of Fig.\,\ref{fig:4} $\varphi_{\text{sqz}}$ and $\varphi_{\text{amp}}$ were slowly scanned, whilst the phase of the BHD’s local oscillator was controlled for amplitude modulation measurements. We used those spectrograms in which the double pass through the squeeze crystal resulted in vacuum noise whilst, at the same time, the signal was amplified to the maximum, with an acceptance window of 5\,\%. This approach resulted in more usable data in a shorter time.
From these the average squeezing value of $(0.9\pm0.17)$\,dB was extracted the same way as before. Based on that the gain of the data was divided by the expectation for a squeezed input alone. This was compared to the expectation based on Eq.\,(\ref{eq:loss-resilience}) with the $1\,\sigma$ band propagating from the squeezing uncertainty and a Monte-Carlo simulation to account for any bias in the analysis. There the signal phase ($\varphi_{\text{sig}}$), random (shot) noise and squeeze value (within its uncertainty) each were randomized and ran through the same analysis.
\\
The implementation in a GW detector requires stable control of these degrees of freedom. As is currently the case in GW detectors, a coherent lock would need to be used for the pump phase \cite{Vahlbruch2006}. For stable control of the signal amplification, maximising the amplification of a phase modulation from the interferometer is a viable option, e.g.~that of a calibration line \cite{Karki2016}.

\subsection*{Theoretical}
The mathematical structure of our model for calculating spectral densities follows Ref.\,\cite{Kimble2001}. Here, we present the case of PDQI for a lossless Michelson interferometer without optical resonators. We derive the complete model in the supplementary material.
We obtain the following input-output relations for the field quadratures $b^q$ and $b^s$
\begin{equation}\label{eq:basicIFO}
    \begin{pmatrix}
        b^q\\
        b^s 
    \end{pmatrix}= 
    e^{2i\phi}
    \begin{pmatrix}
        1 & 0\\
        -\mathcal{K} & 1
    \end{pmatrix}
    \begin{pmatrix}
        a_{\text{in}}^q\\
        a_{\text{in}}^s 
    \end{pmatrix}
    +e^{i\phi}
    \begin{pmatrix}
        0\\
        G_{\text{sig}} 
    \end{pmatrix}.
\end{equation}
Here, $b^s$ and $a^s$ are the output and input in the signal quadrature, respectively; $b^q$ and $a^q$ are the output and input in the quadrature orthogonal to the signal quadrature, respectively; $\mathcal{K}=\mathcal{K}(\Omega)$ is the Kimble factor\,\cite{Kimble2001}; $\phi=\phi(\Omega)$ is the single trip phase of the interferometer. $G_{\text{sig}}$ is the variation of $b^s$ imposed by a GW-signal. It is proportional to the metrics variation $h$ scaled by a constant factor that does note depend on the quantum state of the probing light. The generation of a squeezed state of light can be described by the squeeze matrix
\begin{equation}
    \mathbb{S}[r,\varphi]=
        \begin{pmatrix}
            \cosh r+\cos 2\varphi \sinh r & \sin 2\varphi \sinh r\\ \nonumber
            \sin 2\varphi \sinh r & \cosh r -\cos 2\varphi \sinh r
        \end{pmatrix},
\end{equation}
where $r$ is the squeeze parameter and $\varphi$ the squeeze angle. Applying the squeeze matrix on a vacuum state yields a squeezed state: $a_{\text{SQZ}}=\mathbb{S}[r,\varphi]a_{\text{vac}}$. Reflecting a state of light off a filter cavity causes a rotation of it. In the lossless case, the reflection is described by
\begin{equation}
    \mathbb{F}=
    e^{i \psi}
    \begin{pmatrix}
            \cos \theta(\Omega) & -\sin \theta(\Omega)\\
            \sin \theta(\Omega) & \cos \theta(\Omega)
        \end{pmatrix},
\end{equation}
with the frequency-dependent rotation angle $\theta(\Omega)$, and filter cavity phase $\psi=\psi(\Omega)$. The balanced homodyne detection can be described by a transfer vector. Because we focus on the detection of gravitational waves, the transfer vector is given $\mathbb{H}^{\text{T}} = \left(0, 1\right)$.
To find the condition for the rotation angle of the filter cavity that minimizes the sensitivity of PDQI, we computed the signal quadrature output field
\begin{equation}\label{eq:sig_output}
    \begin{aligned}
        b^s&=\dfrac{e^{2i(\psi+\phi)}}{2}\Biggl[\biggl(2\cos 2\theta(\Omega)+\mathcal{K}\sin 2\theta(\Omega)\biggr)a_{\text{vac}}^s\\
        &-2e^{2r}\cos \theta(\Omega)\biggl(\mathcal{K}\cos \theta(\Omega)-2\sin \theta(\Omega)\biggr)a^q_{\text{vac}}\Biggr]\\
        &+e^{r}e^{i(\psi+\phi)}\cos \theta(\Omega)G_{\text{sig}} \; .
    \end{aligned}
\end{equation}
~\\
 From Eq.\,(\ref{eq:sig_output}), we can see that the cross-correlations created by ponderomotive squeezing vanish if the rotation angle satisfies the condition
\begin{equation}\label{eq:opt_rotation_angle}
    \tan \theta(\Omega)=\dfrac{\mathcal{K}}{2}.
\end{equation}
With Eq.\,(\ref{eq:opt_rotation_angle}) and the spectral density of a vacuum state $S_{\text{vac}}^s=1$, the noise-to-signal ratio of PDQI yields
\begin{equation}
    S_{h}^s=\frac{e^{-2r}\left(4+\mathcal{K}^{2}\right)}{4\eta\vert G_{\text{sig}} \vert^{2}},
\end{equation}
where $\eta=1-l$ is the measurement efficiency. In the same way, we derived the noise-to-signal ratio of the standard frequency-dependent squeeze scheme
\begin{equation} \nonumber
    S_{h,\text{FDS}}^s=\dfrac{1-\eta+\eta e^{-2r}\left(1+\mathcal{K}^{2}\right)}{\eta\vert G_{\text{sig}} \vert^{2}}.
\end{equation}
The improvement of PDQI over frequency-dependent squeezing is as follows
\begin{equation}
    \dfrac{S_{h,\text{FDS}}^s}{S_{h}^s}=\dfrac{4e^{2r}\left(1-\eta\right)+4\eta\left(1+\mathcal{K}^{2}\right)}{4+\mathcal{K}^{2}}.
\end{equation}
In the back-action dominated regime, the improvement due to the additional rotation of the readout quadrature on the output path is approximately given by
\begin{equation}
    \lim_{\mathcal{K}\to \infty}\dfrac{S_{h,\text{FDS}}^s}{S_{h}^s}\approx4\eta.
\end{equation}
In the shot-noise dominated regime, the improvement due to parametric amplification yields
\begin{equation}
    \lim_{\mathcal{K}\to 0}\dfrac{S_{h,\text{FDS}}^s}{S_{h}^s}=e^{2r}\left(1-\eta\right)+\eta.
    \label{eq:expectation}
\end{equation}
\noindent By substituting $\eta = 1-l$ and $\beta  = e^{2r}$ we recover Eq.~\ref{eq:loss-resilience}.

\end{document}